\documentclass[epj-spec]{svjour}
\usepackage{graphicx,amsfonts}
\begin{document}

\newcommand{\bl}[1]{\begin{equation}\label{#1}}
\newcommand{\be}{\begin{equation}}
\newcommand{\ee}{\end{equation}}
\newcommand{\bea}{\begin{eqnarray}}
\newcommand{\eea}{\end{eqnarray}}

\newcommand{\pd}[2]{\frac{\partial{#1}}{\partial{#2}}}
\newcommand{\td}[2]{\frac{\mathrm{d}{#1}}{\mathrm{d}{#2}}}
\newcommand{\rec}[1]{\frac{1}{#1}}
\newcommand{\z}[1]{\left({#1}\right)}
\newcommand{\sz}[1]{\left[{#1}\right]}
\newcommand{\kz}[1]{\left\{{#1}\right\}}
\renewcommand{\sp}{\quad,\quad}
\renewcommand{\v}[1]{\mathbf{#1}}
\newcommand{\m}[1]{\mathrm{#1}}
\renewcommand{\c}[1]{\mathcal{#1}}

\renewcommand{\r}[1]{(\ref{#1})}
\newcommand{\eq}[1]{eq.~(\ref{#1})}
\newcommand{\eqs}[2]{eqs.~(\ref{#1}) and (\ref{#2})}
\newcommand{\eqss}[2]{eqs.~(\ref{#1})--(\ref{#2})}
\newcommand{\Eq}[1]{Eq.~(\ref{#1})}
\newcommand{\Eqs}[2]{Eqs.~(\ref{#1}) and (\ref{#2})}
\newcommand{\Eqss}[2]{Eqs.~(\ref{#1})--(\ref{#2})}


\newcommand{\elte}{ELTE, E{\"o}tv{\"o}s Lor{\'a}nd University, H - 1117 Budapest,
P{\'a}zm{\'a}ny P. s. 1/A, Hungary}
\newcommand{\kfki}{MTA KFKI RMKI, H-1525 Budapest 114, POBox 49, Hungary}

\title{Similar final states from different initial states using new exact solutions
of relativistic hydrodynamics}
\author{M.~Csan\'ad\inst{1}\fnmsep\thanks{\email{csanad@elte.hu}}
   \and M.~I.~Nagy\inst{1}\fnmsep\thanks{\email{nagymarci@rmki.kfki.hu}}
   \and T.~Cs{\"o}rg\H{o}\inst{2}\fnmsep\thanks{\email{csorgo@sunserv.kfki.hu}}}
\institute{\elte\and\kfki}

\abstract{We present exact, analytic and simple solutions of
relativistic perfect fluid hydrodynamics. The solutions allow us
to calculate the rapidity distribution of the particles produced
at the freeze-out, and fit them to the measured rapidity
distribution data. We also give an advanced estimation of the
energy density reached in heavy ion collisions, and an improved
estimation of the life-time of the reaction.}

\maketitle

\section{Introduction}

Since the birth of hydrodynamics, there were many interesting exact solutions found
to the complicated non-linear coupled differential equations of them. In contemporary
research, such as the description of collective properties of high-energy elementary
particle and heavy ion reactions, one has to deal with not only hydrodynamics but
relativistic hydrodynamics. This is an even more complicated topic, and likely this
is the reason why only a few exact solutions exist for relativistic hydrodynamics,
in contrast to the nonrelativistic case.

Exact solutions are important in at least two ways: first, they
can be used to test numerical codes reliably. Second, they provide
an invaluable insight into the details of the evolution of the
matter created in high-energy reactions. After recalling some
presently known solutions, such as the Landau-Khalatnikov solution
and the Hwa-Bjorken solution, we present a class of accelerating
exact and explicit solutions of relativistic hydrodynamics. These
solutions are advantageous compared to the previously mentioned
ones. Then we show how the new solutions can be applied to
describe the evolution of the matter created in high energy
reactions. Our treatment uses simple formulas and takes the
presence of acceleration into account.

This article is based on
refs.~\cite{Csorgo:2006ax,Csorgo:2007ea,Nagy:2007xn}. These
results are extended here with more detailed simulations that
indicate that even for a fixed equation of state, different
initial conditions could lead to the same freeze-out distributions
at mid-rapidity. However, the width of the rapidity distribution
can be utilized to select from among these time evolution
scenarios.

\section{Basic equations}

In this section we recapitulate the equations of relativistic hydrodynamics. We
will use the following notations: $x^\mu=\z{t,\v{r}}$ is the coordinate four-vector,
$\v{r}=\z{r_x,r_y,r_z}$ is the coordinate three-vector. The metric tensor is
$g_{\mu\nu}=diag\z{1,-1,-1,-1}$, the four-velocity field is $u^\mu$,
the normalization is $u^\mu u_\mu=1$. The three-velocity $\v{v}$ is defined
as $u^\mu=\gamma\z{1,\v{v}}$, with $\gamma=\z{1-v^2}^{-1/2}$.
We denote the so-called comoving derivative by $\td{}{t}$, that is,
$\td{}{t}=\pd{}{t}+\v{v}\nabla$. The thermodynamical quantities are
the following: $\varepsilon$ is the energy density, $p$
is the pressure, $w=\varepsilon+p$ is the enthalpy density, $T$
is the temperature, $\sigma$ is the entropy density. When there are
some (conserved or non-conserved) charges present, we denote
them by $n_i$, and the corresponding chemical potentials by
$\mu_i$.

The equations of relativistic hydrodynamics are obtained in the
simplest way by Landau's heuristic argumentation: for perfect
fluid (without viscosity and heat conductivity) the
energy-momentum tensor is
\begin{equation}
T_{\mu\nu}=diag\z{\varepsilon,p,p,p}
\end{equation}
in the local rest frame, so we have
\begin{equation}
T_{\mu\nu}=wu_\mu u_\nu-pg_{\mu\nu}
\end{equation}
in arbitrary frame. Projecting the
energy-momentum conservation law $\partial_\nu T^{\mu\nu}=0$
orthogonal and parallel to $u^\mu$, we obtain the relativistic
Euler equation and the energy conservation equation as \bea
wu^\nu\partial_\nu u^\mu &=& \z{g^{\mu\rho}-u^\mu u^\rho}\partial_\rho p , \label{e:4deul} \\
w\partial_\mu u^\mu      &=& -u^\mu\partial_\mu\varepsilon . \label{e:4denergy}
\eea
The charge conservation equation (for one charge) is
\bl{e:ccons}
\partial_\mu\z{nu^\mu}=0 ,
\ee
but if there are many different particles with the same charge, then this has to
be supplemented by the chemical potentials. For instance, in case of baryonic or
electric charge, particles and antiparticles carry opposite charges, and they
chemical potentials are the same but of opposite sign, so we have
\begin{equation}
\sum_i \mu_i \partial_\mu\z{n_iu^\mu}=0.
\end{equation}

\section{Known solutions}

In this section we summarize some already known exact results: the
renowned Landau-Khalat\-nikov solution, the Hwa-Bjorken solution and
other recent multi-dimensional solutions.

\subsection{The Landau-Khalatnikov solution}

Landau invented relativistic hydrodynamics in the early 1950s and
proposed the application of it to describe high-energy particle
reactions (in that time mostly cosmic ray events). In a
collaboration with I.~M.~Khalatnikov, they found the first exact
solution~\cite{Landau:1953gs,Khalatnikov:1954aa}, which is a 1+1
dimensional implicit solution, with $\varepsilon=3p$ equation of
state (EoS). The Landau-Khalatnikov solution has nice realistic
features: it is an accelerating one, the initial condition
describes a finite piece of matter at rest, then it starts to
expand. Another important feature is that the rapidity
distribution of the particles is approximately Gaussian. However,
the solution itself is very complicated, since the independent
variables --- the time and spatial coordinate --- are given in
terms of horrendous integral formulas involving Bessel functions
integrated over the temperature and the fluid rapidity
``variables". Thus we don't quote the whole result here.

\subsection{The Hwa-Bjorken solution}

The Hwa-Bjorken solution~\cite{Hwa:1974gn,Bjorken:1982qr} is a 1+1
dimensional, expanding accelerationless solution. It is simple and
explicit, this is its main advantage compared to the
Landau-Khalatnikov solution. It is written down in the easiest way
in Rindler-coordinates $\tau$ and $\eta$, which are defined by
\begin{equation}\label{e:Rin}
t=\tau\cosh\eta \sp r=\tau\sinh\eta
\end{equation}
with $r$ being the spatial coordinate. The Hwa-Bjorken solution is given by the
velocity and entropy profiles
\begin{equation}
v=\tanh\eta=\frac{r}{t} \sp \sigma=\sigma_0\frac{\tau_0}{\tau} ,
\end{equation}
where the subscript $_0$ refers to the initial condition. It works
for $\epsilon + B = \kappa (p + B)$, for arbitrary values of
$\kappa$ and $B$. This boost-invariant solution results in a
constant rapidity distribution of the particles. This is a rough
prediction clearly not valid at the present experimental
situations. However, Bjorken has given a simple estimate of the
initial energy density of the reaction based on this solution,
this is why it became so renowned. In the next section we present
such solutions that are simple and explicit (as the Hwa-Bjorken
solution) and are accelerating, and yield realistic rapidity
distributions (as the Landau-Khalatnikov solution).

\section{New simple analytic solutions}

The Landau-Khalatnikov solution contains acceleration, and yields a finite,
realistic, Gaussian-like rapidity distribution. On the other hand, the
Hwa-Bjorken solution is very simple and easy to handle, this is why it
became so important, although the accelerationless, boost-invariant flow
profile and the constant rapidity distribution obviously does not agree
with the results of real high-energy experiments.

We found such new, analytic, explicit and simple solutions, which
do not lack acceleration, and yield finite, realistic rapidity
distributions~\cite{Csorgo:2006ax,Csorgo:2007ea}. We found
solutions for 1+1 dimensions, and also spherical flows in
arbitrary number of spatial dimensions ($d$ will denote their
number). For presenting the solutions we use Rindler-coordinates,
which are defined by \eq{e:Rin} for $t>r$, that is ,,inside the
lightcone''. (From now on $t$ is the time and $r$ is the radial
coordinate in the 1+d dimensional case, and the single spatial
coordinate in the 1+1 dimensional case.)

For a discussion of the mathematical derivations in greater
detail, apart from Refs.~\cite{Csorgo:2006ax,Csorgo:2007ea}, we
recommend Ref.~\cite{Nagy:2007xn}. Here we present only the main
results. The velocity and the pressure is given by
\begin{equation}
v=\tanh(\lambda \eta)
\sp p=p_0\z{\frac{\tau_0}{\tau}}^{\lambda d \frac{\kappa+1}{\kappa}}
\rec{\cosh^{\phi_{\lambda}(d-1)}\frac{\eta}{2}} .
\end{equation}
The value of the constants $\lambda$, $\phi_{\lambda}$, $d$ and
$\kappa$ are constrained: different set of values yield different
solutions. Table~\ref{t:sol} shows the possible cases, every row
of the table being a solution. In the following we discuss them
separately.
\begin{table}
\begin{center}
\begin{tabular}{|c|c|c|c|c|}
  \hline
  Case & $\lambda$ &       $d$             & $\kappa$ &         $\phi_{\lambda}$            \\ \hline
  a.)  & $1$             & $\in\mathbb{R}$ & $\in\mathbb{R}$  & $0$                         \\
  b.)  & $2$             & $\in\mathbb{R}$ & $d$              & $0$                         \\
  c.)  & $\rec{2}$       & $\in\mathbb{R}$ & $1$              & $\frac{\kappa + 1}{\kappa}$ \\
  d.)  & $\frac{3}{2}$   & $\in\mathbb{R}$ & $\frac{4d-1}{3}$ & $\frac{\kappa + 1}{\kappa}$ \\
  e.)  & $\in\mathbb{R}$ & $1$             & $1$              & $0$                         \\  \hline
\end{tabular}
\end{center}
\caption{The new family of solutions.}\label{t:sol}
\end{table}
\begin{itemize}

\item Case a)\@ isn't a new solution, it is just the well-known
Hwa-Bjorken solution in 1+1 dimensions, and the only recently
discovered Buda-Lund type of
solutions~\cite{Csorgo:2003rt,Csorgo:2003ry} in 1+d dimensions. We
quoted it for the sake of completeness, since this solution is
also a member of the class. Apart from this case, all the other
solutions possess non-vanishing acceleration.

\item Case b)\@ was found first, with other methods. It describes
spherical expansion in $d$ dimensions. The EoS is constrained in a
way that $\kappa$ must be equal to $d$. (For instance, in case of
three-dimensional expansion $\kappa=3$, which is the EoS of an
ultra-relativistic ideal photon gas.) It is interesting to
calculate the equation of trajectories $R(t)$ from the velocity
field in Minkowskian coordinates:
\begin{equation}
v=\frac{2tr}{r^2+t^2}
\quad\Rightarrow\quad R(t)=\rec{a_0}(\sqrt{1+(a_0 t)^2}+1)
\end{equation}
with $a_0=\frac{2r_0}{\left|r_0^2-t_0^2 \right|}$. So the fluid
elements have constant $a_0$ acceleration in the local rest frame.

\item Case $c)$ and $d)$ were found first by T.~S.~Bir\'o in 1+3
dimensions~\cite{Biro:2007pr}, we generalized them to arbitrary
number of spatial dimensions. These solutions are also
accelerating, and since $\phi_{\lambda}\neq 0$, if $d\neq 1$, the pressure
field explicitly depends on $\eta$, so the solution is finite in
$\eta$.

\item Case $e)$ has a remarkably general velocity field: the
$\lambda$ parameter can be arbitrary. We call $\lambda$
acceleration parameter, because it somehow governs the
acceleration of the flow. On the other hand, this solution works
only for $d=1$ and $\kappa=1$, which is obviously a drawback.
Nevertheless, this solution can be considered as a ,,smooth
extrapolation'' between the previous cases. In the next section we
shall use this solution to calculate the rapidity distribution,
and see that the width of it is controlled by the value of
$\lambda$. Hence in principle, the value of the parameter
$\lambda$ of the hydrodynamical solution can be obtained from
measurements.
\end{itemize}
For a more thorough review of these solutions we recommend
ref.~\cite{Nagy:2007xn}.

For illustration, in Fig.~\ref{f:anim} we plot the spatial
temperature distribution at various points of the evolution of a
collision for different $\lambda$ values. All the plotted
solutions go to the same final temperature at mid-rapidity
($\eta=0$), although their initial temperature and acceleration
parameter $\lambda$ are different. We shall see in the next
section, that solutions with different acceleration parameter
$\lambda$ have different rapidity distributions, and the
acceleration parameter can be determined from the measurement of
the widths of the rapidity distribution.

\begin{figure}
\includegraphics[width=0.497\linewidth]{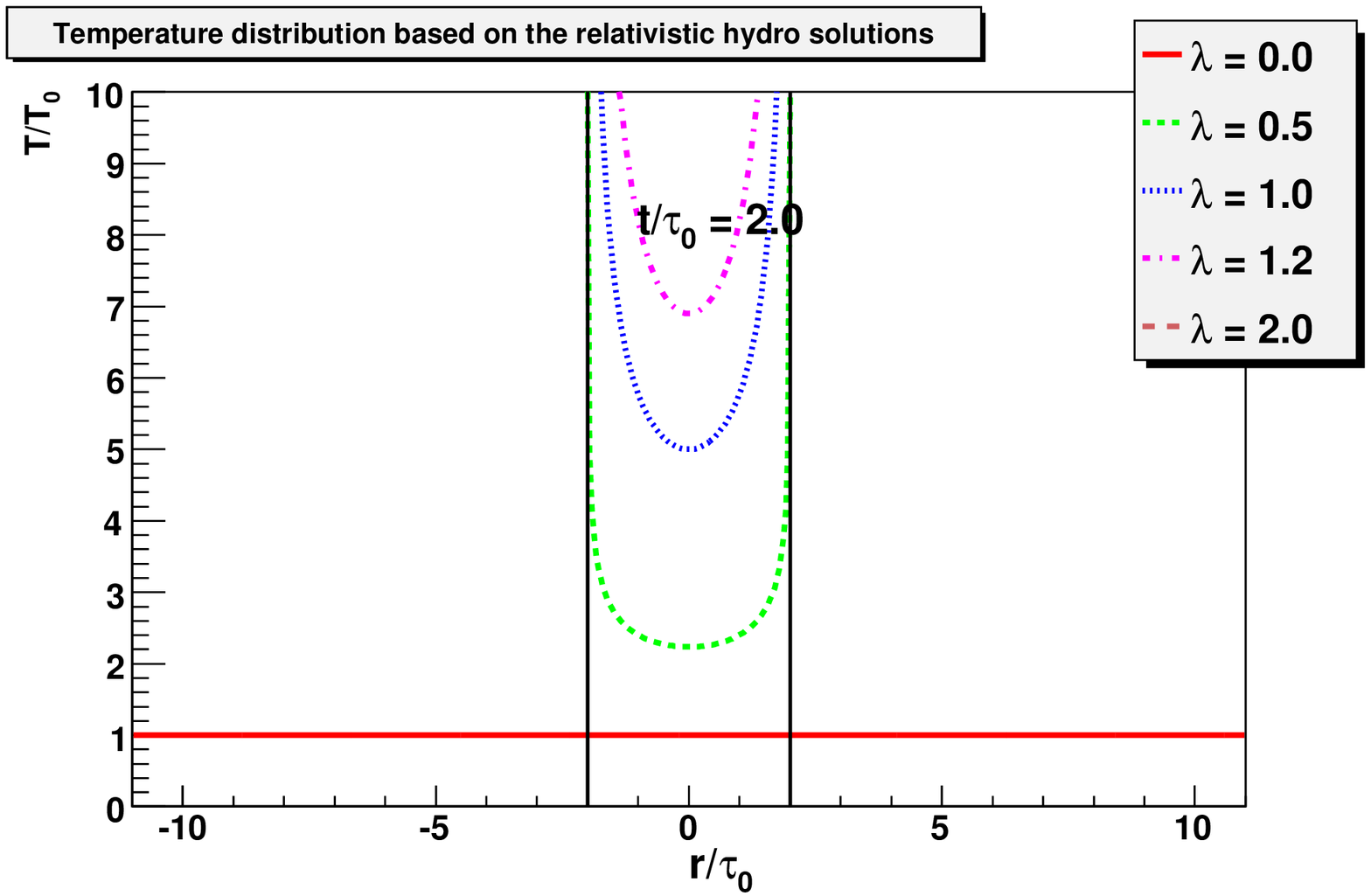}
\includegraphics[width=0.497\linewidth]{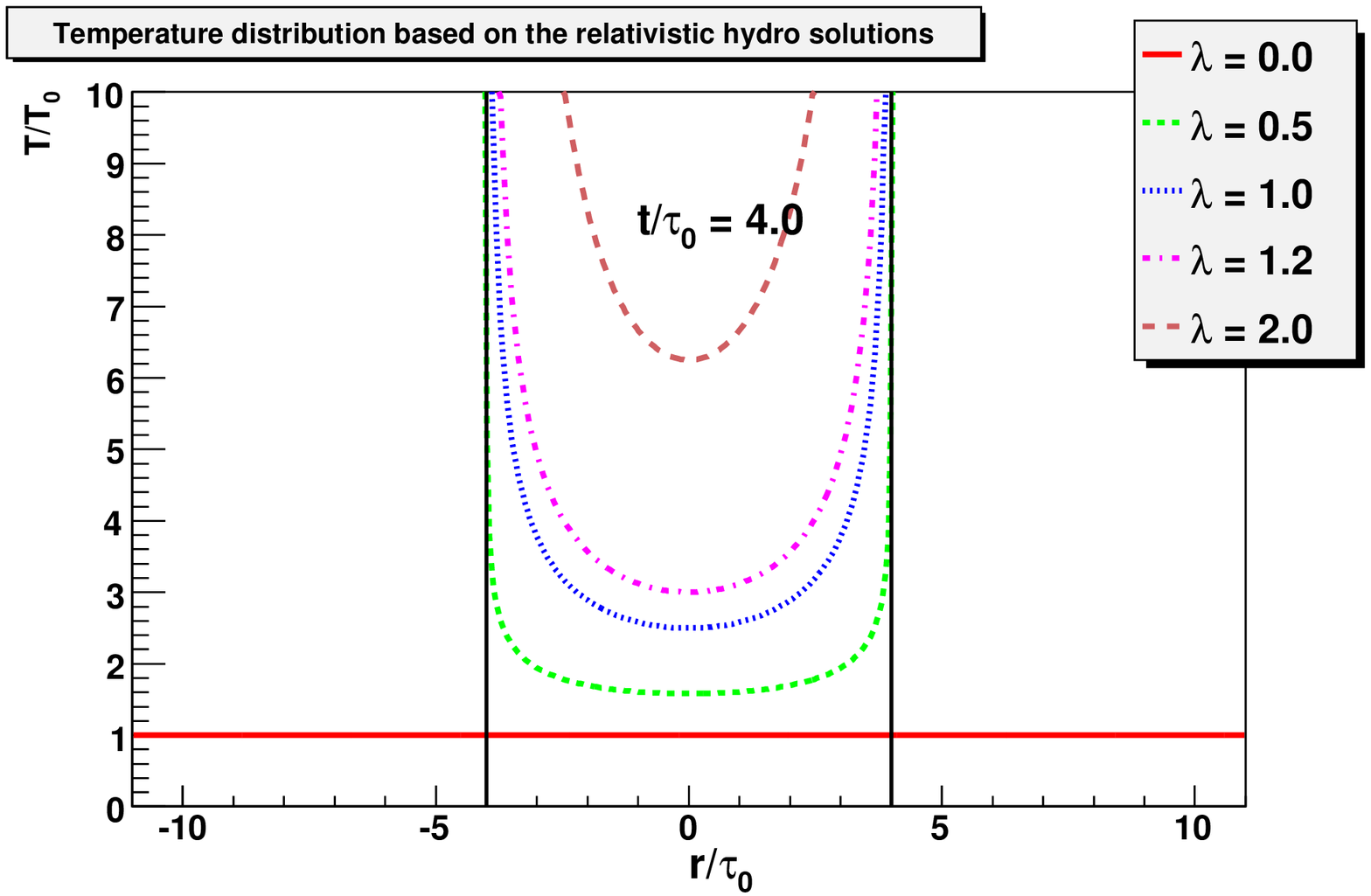}
\includegraphics[width=0.497\linewidth]{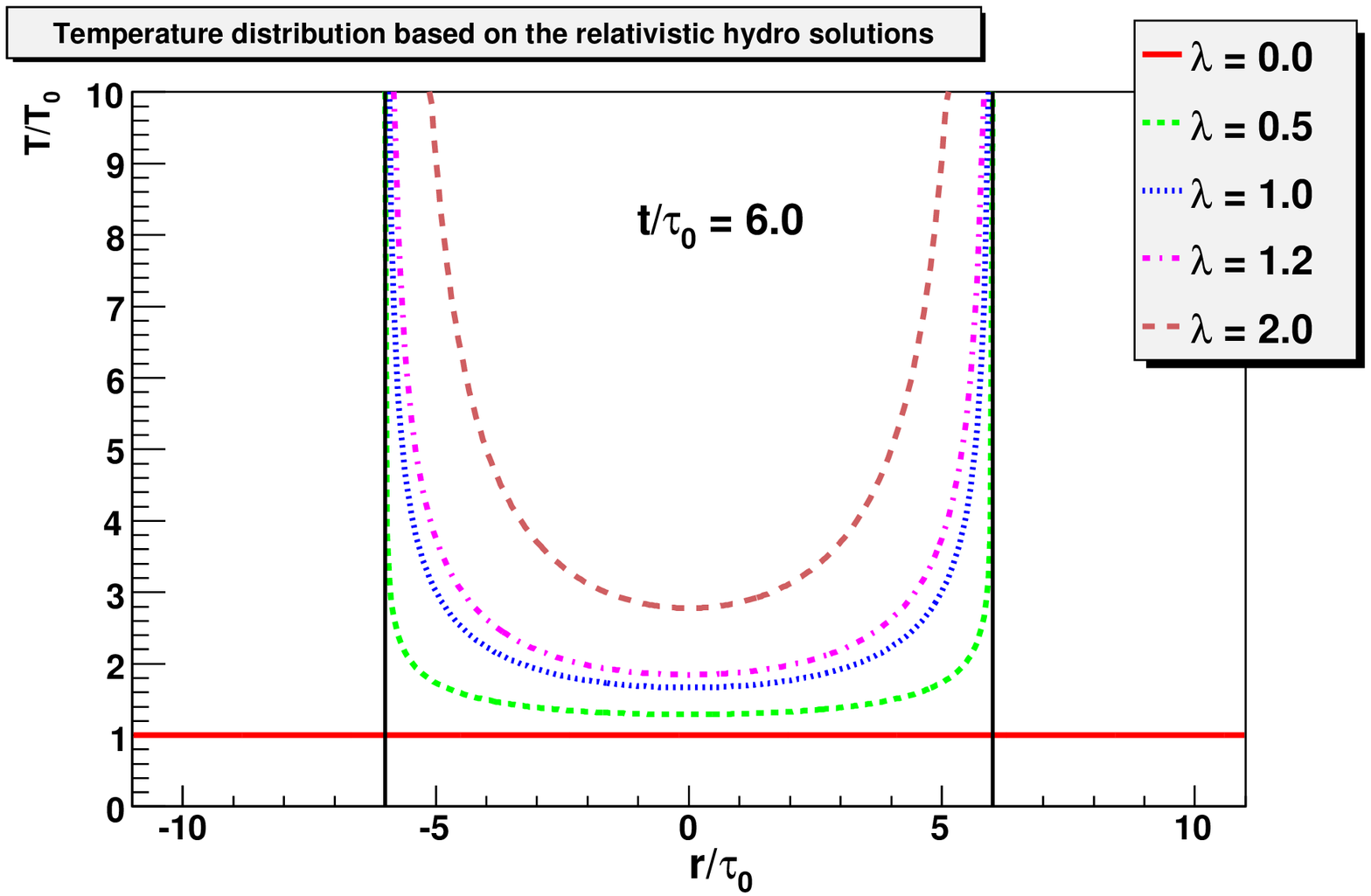}
\includegraphics[width=0.497\linewidth]{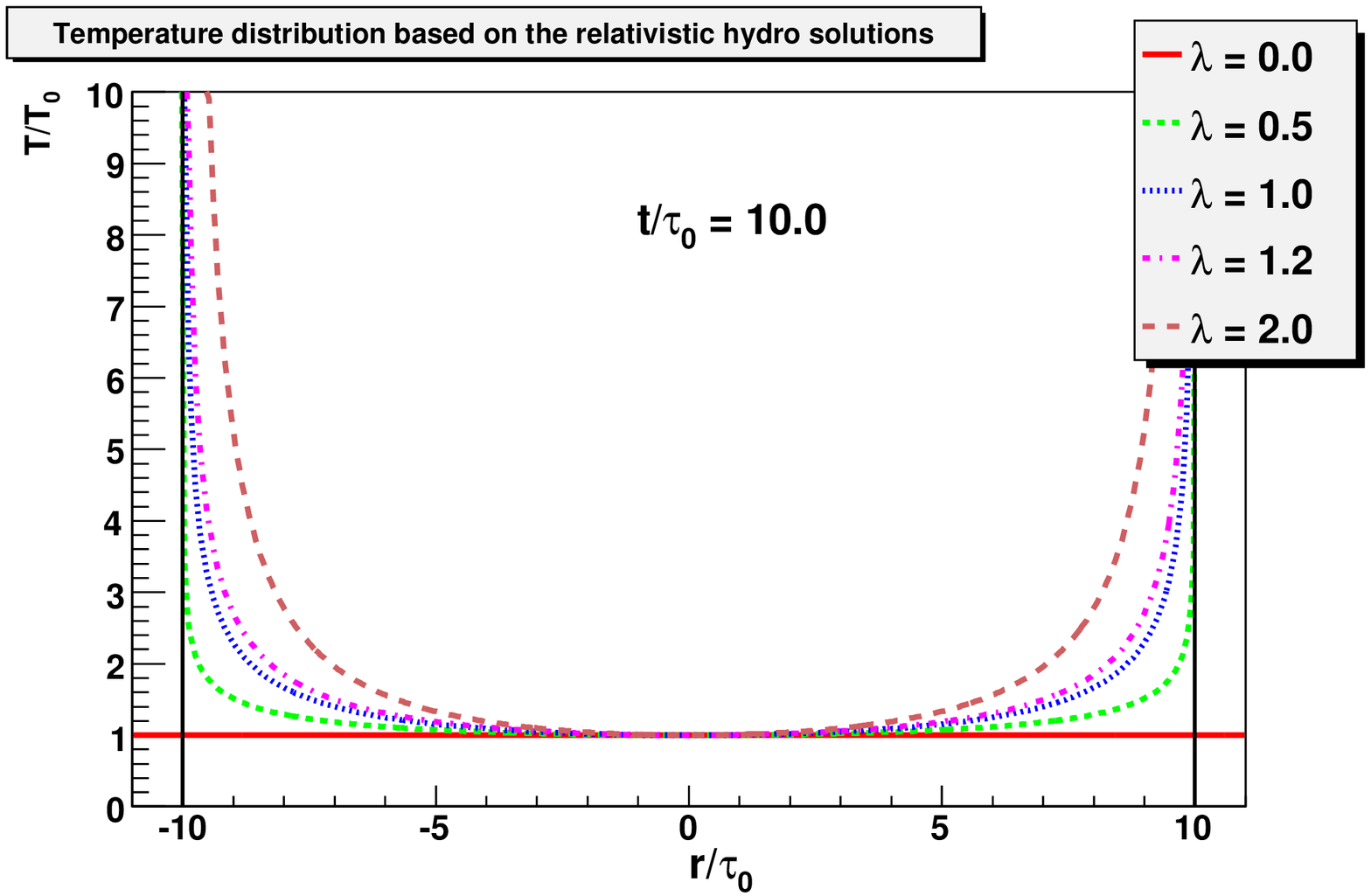}
\caption{\label{f:anim}(Color online) These plots show the
temperature distribution of different one dimensional solutions
(case e. of Table~\ref{t:sol}, with different $\lambda$
parameters) with different initial temperature. All the plotted
solutions go to the same final temperature at mid-rapidity
($\eta=0$). The $\lambda=1$ solution corresponds to the
Hwa-Bjorken solution. All the $\lambda > 0$ solutions are shown
within the light-cone (vertical solid black lines). The horizontal
solid red line indicates the $\lambda = 0$ case, corresponding to
a static, infinite, homogeneous medium with a constant
temperature, and its domain extends beyond the light-cone, while
for $\lambda > 0$ we plot the solutions at $r<t$ only.}
\end{figure}

\section{Rapidity distribution}

We calculated the rapidity distribution, $\td{n}{y}$ of the
particles numerically and (approximately) analytically as well. We
used the solutions discussed in the previous section as case
$e.)$, that is, where the $\lambda$ parameter is arbitrary. We
assumed that at a certain freeze-out massive particles appear. (In
the figures of this paper we used pions with $m=140$ MeV, but
other particles can be used as well -- and their contribution can
be added to the one of pions). We have chosen the freeze-out
condition as follows: the temperature in $\eta=0$ should reach a
given $T_f$ value (subscript $_f$ means freeze-out) and the
freeze-out hypersurface should be pseudo-orthogonal to the
four-velocity field $u_\mu(x)$. The equation of this hypersurface
is \be
\z{\frac{\tau_f}{\tau}}^{\lambda-1}\cosh\z{(\lambda-1)\eta}=1 .
\ee The details of the calculation of  $\td{n}{y}$ is found in
refs.~\cite{Csorgo:2006ax,Nagy:2007xn}. We only quote the result
here: using a saddle-point integration in $\eta$, for
$\lambda>1/2$, $m/T_f \gg 1$ and $\nu_{\sigma}(s)=1$ we got
\begin{equation}\label{e:dndy-approx}
\td{n}{y}\approx\td{n}{y}\Big{|}_{y=0}
                       \cosh^{-\frac{\alpha}{2}-1}\z{\frac{y}{\alpha}}
                       e^{-\frac{m}{T_f}\sz{\cosh^\alpha\z{\frac{y}{\alpha}}-1}} ,
\end{equation}
where $\alpha=\frac{2\lambda-1}{\lambda-1}$. Some typical cases
are plotted in Fig.~\ref{f:rapdist}. This figure is just an
illustration, the parameter values on this figure are not
realistic ones.
\begin{figure}
\includegraphics[width=0.497\linewidth]{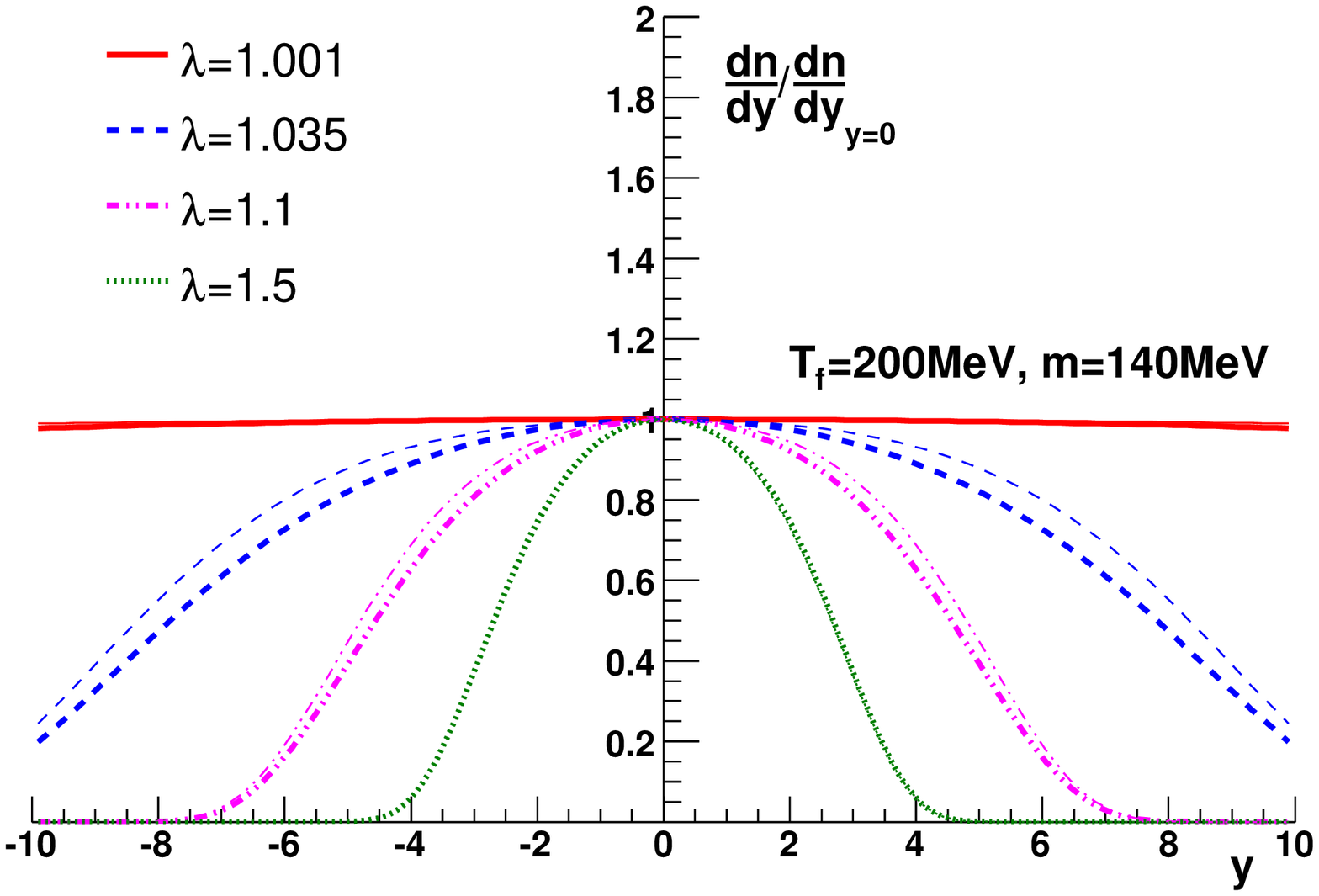}
\includegraphics[width=0.497\linewidth]{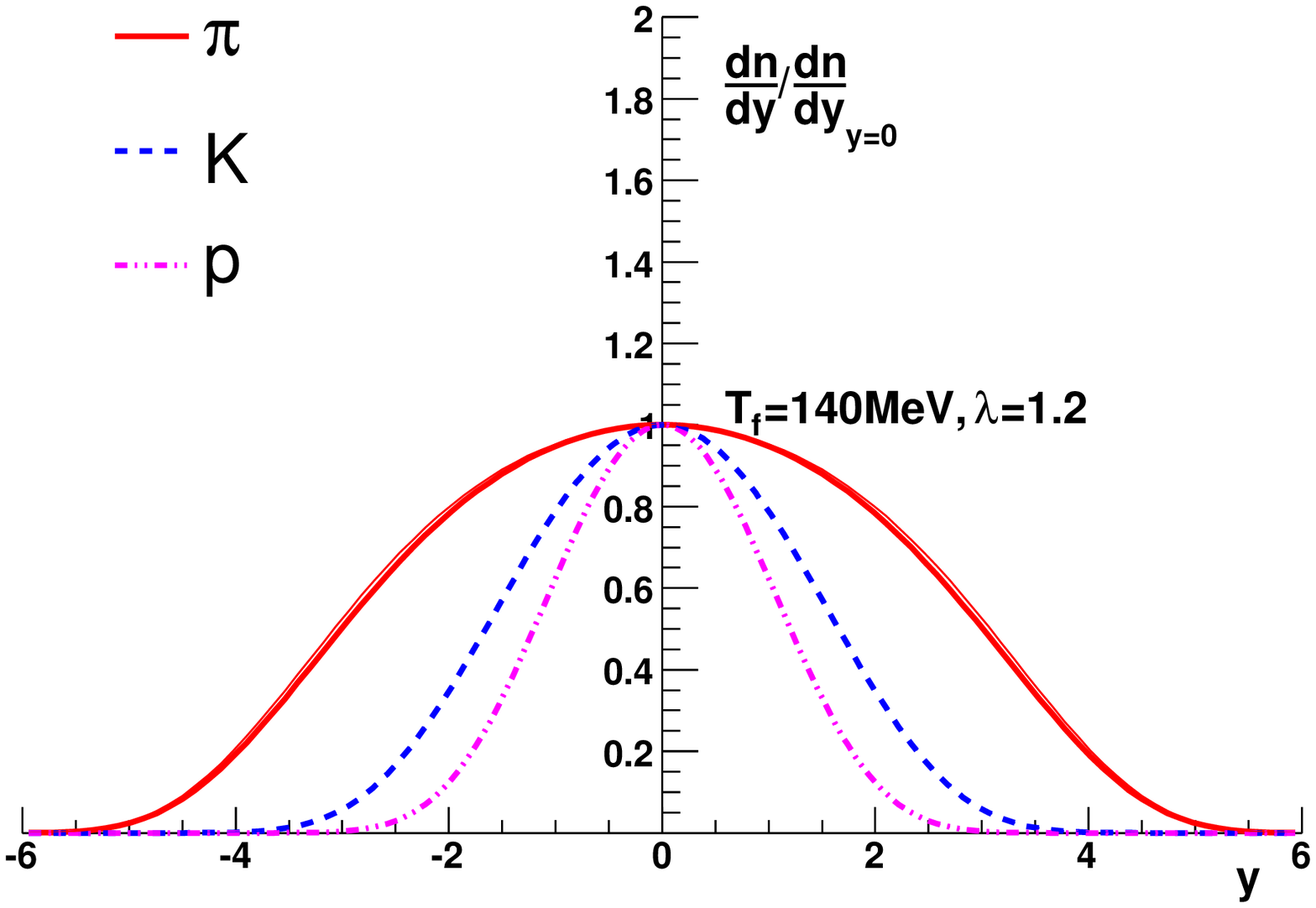}
\caption{\label{f:rapdist}(Color online) Normalized rapidity
distributions from the new solutions in 1+1 dimensions -- case e.)
of Table~\ref{t:sol} -- for various $\lambda$, $T_f$ and $m$
values. Thick lines show the result of numerical integration, thin
lines the analytic approximation from \eq{e:dndy-approx}. For
$\lambda>1$ and not too big $T_f$ it can be used within about 10
\% error. }
\end{figure}

\section{Energy density estimation}

Now as we have an analytic approximation for the rapidity
distribution, we are able to fit it to real experimental data and
extract the $\lambda$ acceleration parameter from them. Since our
treatment includes acceleration effects, the new solutions provide
a more realistic picture of the initial period of high energy
reactions. In this section we show how this can be used for
improving the famous energy density estimation made by Bjorken. It
is clear that initial energy density is a quantity of crucial
importance when one wants to interpret the conclusions drawn from
high energy experiments.

We follow Bjorken's method~\cite{Bjorken:1982qr} and modify it when acceleration
effects become important. Let us focus on a thin transverse piece of the produced
matter at mid-rapidity, as seen on by Fig. 2 of ref.~\cite{Bjorken:1982qr}. The
radius $R$ of this slab is estimated by the radius of the colliding hadrons or nuclei:
$R=1.18A^{1/3}$fm. The volume is $dV=(R^2\pi)\tau\m{d}\eta$, where $\tau$ is the
proper time of observation and $\m{d}\eta$ is the space-time rapidity extent of
the slab. The energy content $dE =\langle m_t\rangle dn$, where $\langle m_t\rangle$
is the average transverse mass at mid-rapidity, so similarly to Bjorken, the
initial energy density is
\bl{e:Bjorken}
    \varepsilon_0 = \frac{\langle m_t\rangle}{(R^2 \pi)\tau_0}\frac{dn}{d\eta_0} .
\ee Here $\tau_0$ is the proper-time of thermalization, estimated
by Bjorken as $\tau_0\approx 1$fm. For accelerationless,
boost-invariant Hwa-Bjorken flows $\eta_0=\eta_f=y$, however, for
accelerating solutions one has to apply a correction of
\begin{equation}
\pd{y}{\eta_f}\pd{\eta_f}{\eta_0}=\z{2\lambda-1}\z{\tau_f/\tau_0}^{\lambda-1}
\end{equation}
(see Fig.~\ref{f:corr}). These two factors contain the
acceleration effects on the energy density estimation, see
ref.~\cite{Nagy:2007xn} for details.

\begin{figure}
\begin{center}
\includegraphics[width=0.95\linewidth]{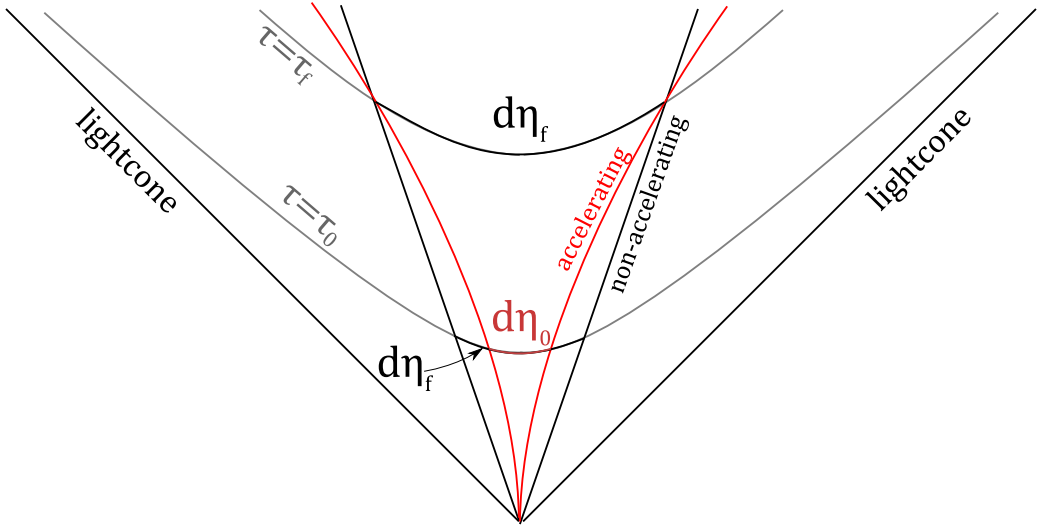}
\end{center}
\caption{\label{f:corr}(Color online) This figure shows that if
there is no acceleration, $\eta_0=\eta_f$, but for the
accelerating case, a correction factor has to be applied.}
\end{figure}

So the initial energy density $\varepsilon_0$ can be accessed by an
advanced estimation $\varepsilon_c$ as
\bl{e:ncscs}
\frac{\varepsilon_c}{\varepsilon_{Bj}}=\z{2\lambda-1}\z{\frac{\tau_f}{\tau_0}}^{\lambda-1}
\sp \varepsilon_{Bj}=\frac{\langle m_t\rangle}{(R^2\pi)\tau_0}\frac{dn}{dy} .
\ee
Here $\varepsilon_{Bj}$ is the Bjorken estimation, which is recovered
if $\td{n}{y}$ is flat (i.e. $\lambda=1$), but if $\lambda>1$,
i.e. the flow is accelerating, both correction factors are greater
than $1$, so $\varepsilon_0$ is \emph{under-estimated} by the Bjorken formula.
Fig.~\ref{f:estim} shows our fits to BRAHMS $dn/dy$
data~\cite{Bearden:2004yx}. From these fits we have found
$\lambda=1.18\pm 0.01$.

Using the Bjorken estimate of $\varepsilon_{Bj} = 5$ GeV/fm$^3$ as given in
ref.~\cite{BRAHMS-White}, and $\tau_f/\tau_0=8\pm 2$ fm/c, we find
an initial energy density of $\varepsilon_c = (2.0\pm
0.1)\varepsilon_{Bj}=10.0\pm 0.5$ GeV/fm$^3$.

\begin{figure}
\includegraphics[width=0.497\linewidth]{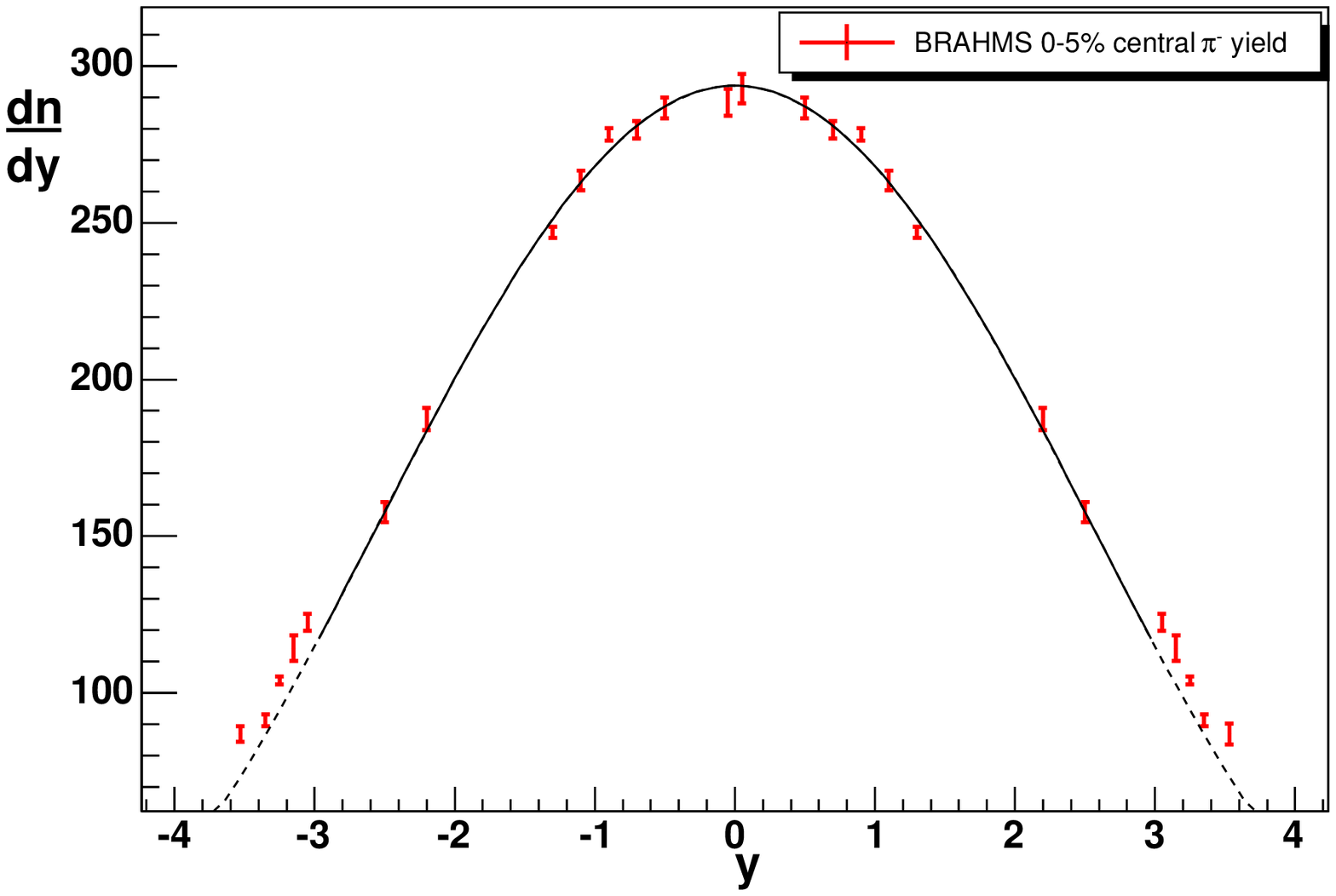}
\includegraphics[width=0.497\linewidth]{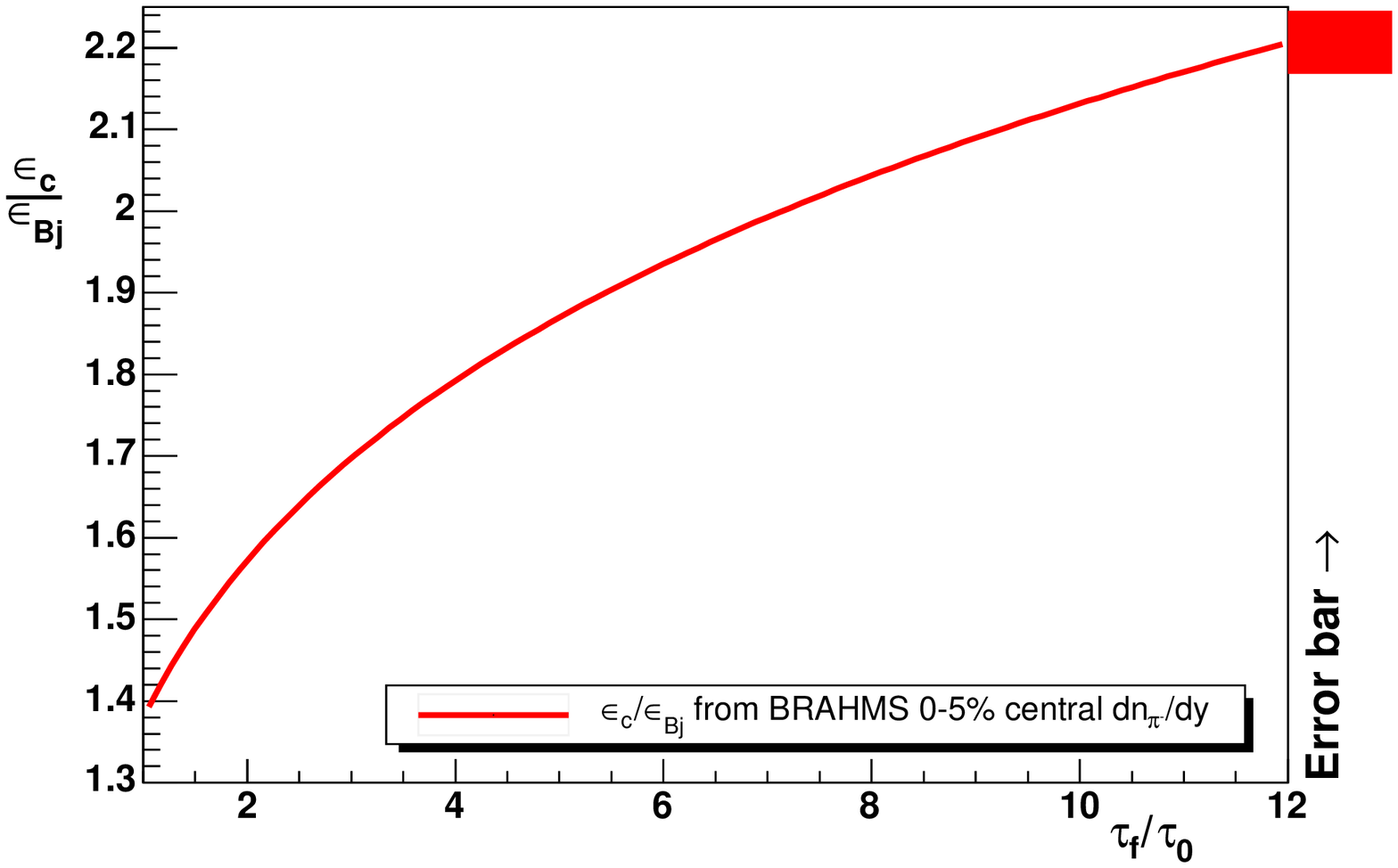}
\caption{\label{f:estim}(Color online) Left panel: $dn/dy$ data of
negative pions, as measured by the BRAHMS
collaboration~\cite{Bearden:2004yx} in central (0-5\%) Au+Au
collisions at $\sqrt{s_{NN}}=200$ GeV, fitted with
\eq{e:dndy-approx} (1+3 dimensional case). The fit range was $-3 <
y < 3$, to exclude target and projectile rapidity region, CL = 0.6
\%. Right panel: $\varepsilon_c/\varepsilon_{Bj}$ ratio as a
function of $\tau_f/\tau_0$.}
\end{figure}

\section{Life-time estimation}

As another application, let us focus on the life-time of a high energy reaction.
For a Hwa-Bjorken type of accelerationless, coasting  longitudinal flow,
Sinyukov and Makhlin~\cite{Makhlin:1987gm} determined the longitudinal
length of homogeneity as
\bl{e:SM_Rlong}
R_{long} = \sqrt{\frac{T_f}{m_t}}\tau_{Bj} .
\ee
Here $m_t$ is the transverse mass and $\tau_{Bj}$ is the (Bjorken)
freeze-out time. This result provides a means to determine the life-time
of the reaction, if one simply identifies it with $\tau_{Bj}$.
However, if the flow is accelerating, the case is a little bit more complicated:
the estimated origin of the trajectories is shifted back in proper-time, so
the life-time of the reaction is under-estimated by $\tau_{Bj}$. From our
solutions we have (for a broad but finite rapidity distribution,
so that the saddle-point approximation can be used):
\bl{Rlong-c}
R_{long}=\sqrt{\frac{T_f}{m_t}}\frac{\tau_c}{\lambda}
    \quad\Rightarrow\quad
    \tau_{c} = \lambda \tau_{Bj} .
\ee
Thus the new estimation of the life-time, $\tau_c$ contains a $\lambda$
multiplication factor. BRAHMS rapidity distributions in Fig.~\ref{f:estim}
yield $\lambda=1.18\pm 0.01$, so they imply a 18 $\pm$1 \% increase in
the estimated life-time of the reaction.

\section{Summary}

We presented a class of simple solutions of relativistic perfect
fluid hydrodynamics. Because the new solutions are accelerating,
but still explicit and exact, they can be utilized in the
description of high energy reactions in a more reliable way than
either the Hwa-Bjorken of the Landau-Khalatnikov solution. We have
shown, that even for a fixed equation of state, different initial
conditions could lead to the same freeze-out distributions at
mid-rapidity. However, these solutions of different acceleration
parameter $\lambda$ have different width of their rapidity
distributions, and the acceleration parameter can thus be
determined from the measurement of the widths of the rapidity
distribution. We have calculated the rapidity distribution of the
produced particles, and showed two possible applications: and
advanced estimation of the initial energy density of the high
energy reaction and an advanced life-time estimation. We fitted
the analytic approximation of the rapidity distribution to
experimental data (measured by the BRAHMS collaboration at RHIC)
and determined the $\lambda$ acceleration parameter of the flow.
From it we concluded that the energy density estimations based on
Bjorken's method has to be corrected by a factor of $~2$. We also
obtained a correction to the life-time measurements: a $\sim 20$\%
increase is due to the presence of acceleration in the initial
evolution of the matter. Thus the effects of work, done even by
the fluid cell at mid-rapidity, can be determined from the
measurement of the widht of the  rapidity distribution and this
information is important for an advanced estimation of initial
energy densities and the life-time of the rapidity distribution.

\end{document}